# Allometry and growth of eight tree taxa in United Kingdom woodlands

Matthew R. Evans[1], Aristides Moustakas[1], Gregory Carey[1], Yadvinder Malhi[2], Nathalie Butt[3], Sue Benham[4], Denise Pallett[5], Stefanie Schäfer[5]

[1]School of Biological and Chemical Sciences, Queen Mary University of London, Mile End Road, London, E1 4NS, United Kingdom

[2]Environmental Change Institute, School of Geography and the Environment, University of Oxford, OX1 3QY, UNITED KINGDOM

[3]ARC Centre of Excellence for Environmental Decisions, School of Biological Sciences, The University of Queensland, St. Lucia, 4072, Australia

[4]Centre for Ecosystem, Society and Biosecurity, Forest Research, Alice Holt Lodge, Wrecclesham, Farnham, Surrey, GU10 4LH, United Kingdom

[5]CEH-Wallingford, Maclean Building, Benson Lane, Crowmarsh, Gifford, Wallingford, Oxfordshire, OX10 8BB, United Kingdom

Corresponding author: Matthew Evans (m.evans@qmul.ac.uk)

## Abstract

As part of a project to develop predictive ecosystem models of United Kingdom woodlands we have collated data from two United Kingdom woodlands - Wytham Woods and Alice Holt. Here we present data from 582 individual trees of eight taxa in the form of summary variables relating to the allometric relationships between trunk diameter, height, crown height, crown radius and trunk radial growth rate to the tree's light environment and diameter at breast height. In addition the raw data files containing the variables from which the summary data were obtained. Large sample sizes with longitudinal data spanning 22 years make these datasets useful for future studies concerned with the way trees change in size and shape over their life-span.

## Background & Summary

Prediction is a basic, possibly defining, feature of scientific disciplines [1]. To develop ecological models that are capable of being projected into the future, possibly into novel conditions outside the parameter space within which the data were collected, process-based models are required. Such process-based models are extremely demanding of data, as there are often many interacting processes each requiring parameterisation [2,3]. For long-lived species, such as trees, parameterisation is especially demanding as most processes occur slowly, and so require long-term datasets to ensure that robust estimates of the relevant rates can be obtained [4]. It is rare that datasets exist for the purposes of creating such models, and so data, the collection of which was originally motivated by some other purpose, usually need to be identified and processed in a manner that makes them suitable for inclusion in such models. At present in ecology, prediction is attempted relatively rarely [5,6] and for example the recent United Kingdom National Ecosystem Assessment struggled to find suitable models or empirical examples on which to base its scenarios of likely future states of ecosystems [7].

We are developing predictive ecosystem models initially with the intention of providing projections of the future state of United Kingdom woodlands. As our underlying computational model we have implemented SORTIE – an established forest model [8]. We chose SORTIE over the many competing models because it is conceptually simple (based on trees competing for one resource, i.e. light), it is based on ecological information that can be parameterised from field data, it has been extensively and successfully used in North America [8-11] and New Zealand [12-14], and it is individual-based, which allows for us to plan for coupling between trophic levels more easily than if individuals were aggregated. In SORTIE, trees compete for light by intercepting incident sunlight and modifying the light environment beneath their crown. Sapling growth depends on their light environment while adult growth depends on their size. For adult trees, traits (height, crown height and radius) vary with diameter at breast height (DBH); while for saplings, traits vary with diameter at 10cm above ground level. We have parameterised these functions by collating three datasets, and by collecting data specifically for this project where they did not previously exist. Here we make available these data and the summary variables for the eight commonest tree taxa (Table 1).

This information is most obviously of utility for those, who like ourselves, are planning to use individual-based models of trees, and who may be interested in the allometry and growth of the taxa included here (Table 1). However, allometric relationships such as these are extremely important in understanding the biology of the species concerned [15,16] and so will be of interest to those with more fundamental ecological interests. Similarly practitioners, e.g. foresters, may find these data of use if they wish to understand how timber production changes as trees grow. DBH has long been the measurement of choice among foresters – for good reason as it is both straightforward to measure and interpret in terms of timber volume [17]. The data presented here allow estimation of other aspects of tree size and shape from DBH.

Since 1992 the Environmental Change Network has measured DBH and height of focal trees at two woodland sites – Wytham Woods (Oxfordshire) and Alice Holt (Surrey) using standard protocols [18]. The DBH of an additional set of trees was also measured in Wytham Woods in 2008 and 2010 (by two of us – Malhi and Butt). We have collated these data and combined them into a single dataset, which we have supplemented with data on the crown height and crown radius of the adult trees, diameter at 10 cm above ground level, and on the local light environment of saplings. The workflow used to generate the output is shown in Fig. 1.

## Methods

### Study sites

Data were collected from two United Kingdom woodlands – Wytham Woods and Alice Holt. Wytham Woods (51˚46'N, 1˚20'W, UK National Grid: SP 46 08) has been a research site (owned and managed by Oxford University) since the 1940s. It is approximately 400 Ha in extent, ranging in height from 60-165 m above sea level. The site has been extensively managed, mainly by coppicing, although this has not been conducted since the early 20th Century. There are regions of ancient woodland, secondary woodland and plantation; only the plantation areas are still managed today. Alice Holt (51˚10'N, 0˚50'W, UK National Grid: SU80 42) is in northern Hampshire and managed by the Forest Commission. The entire site is about 850 Ha, the majority planted with Corsican pine (*Pinus nigra*), but 140 Ha of old-growth oak (*Quercus robur*) woodland remain, in which the data used here were collected. The site varies in altitude from 70-125 m. These two sites are the two woodland sites in the Environmental Change Network (ECN) in the United Kingdom.

**Data collection**

Three datasets have been collated here:

- Environmental Change Network (Wytham Woods) (ECN-W), the ECN has monitored a fixed set of sites in Wytham Woods since 1992. In this dataset we have data on 250 individual trees in 41 plots that have had DBH measured on seven occasions (1993, 1996, 1999, 2002, 2005, 2008, 2012), and height measured three times (1993, 2002, 2012)
- Environmental Change Network (Alice Holt) (ECN-AH), the ECN has monitored a fixed set of sites in Alice Holt since 1994. In this dataset we have data on 216 individual trees in 51 plots that have DBH measured seven times (1994, 1997, 2000, 2002, 2005, 2008, 2011). Height was measured on the same trees on three occasions (1994, 2002, 2011). Another set of 56 trees on nine plots had DBH measured five times (2004, 2005, 2007, 2012 and 2013), and height measured twice (2004 and 2012)
- Oxford University plot (OXF), two of the authors (Malhi and Butt) have established an 18 Ha plot containing about 20,000 individual trees, which have had DBH measured on two occasions (2008 and 2010)

Since 2011 three of the authors (Evans, Moustakas and Carey) have supplemented these three datasets by:

- Measuring diameter at 10 cm above ground level on all saplings (defined here as trees with DBH < 10cm) in ECN-W, ECH-AH, and a sample of 88 from OXF
- Measuring the light environment around all saplings in ECN-W, ECH-AH and a sample of 88 from OXF
- Measuring the height of a sample of 88 saplings from OXF
- Measuring crown radius and crown height for all adults in ECN-W and ECH-AH
- Measuring canopy openness for a sample of 165 trees in Wytham Woods

It would have been desirable to have estimates of the age of trees in the datasets. Unfortunately none of trees in datasets have been cored to determine tree age. In a separate publication we have estimated tree mortality for the same taxa as are included here using the ECN-W dataset through the application of a Cormack Jolly Seber model [19].

**Measurement methods**

*Diameter at Breast Height (DBH)*

DBH is a measurement that is routinely included in the datasets collated here. The three datasets ECN-W, ECN-AH and OXF include a measurement of DBH which is taken following standardised methods, by measuring trunk circumference using a tape to the nearest 0.1 cm at 1.3 m above ground level [18]. To ensure that DBH was measured at the same point on subsequent surveys trees were marked with paint at the point at which DBH was measured.

*Growth*

Mean growth rates of individual trees were estimated by taking a series of DBH measurements and subtracting the measurement at time point t from the measurement at t+1 to calculate the change in DBH between the two time points and then to divide this value by the number of years between the two time points. If for any tree there were more than two measurements, the values were averaged to produce a single value per tree.

*Height (H)*

Tree height is measured in the two ECN datasets, and was measured by Evans and Moustakas for a number of further trees, as described above. Height is measured by ECN using a hypsometer to the nearest 0.5 m at Wytham Woods following [18], and using a laser Vertex (Haglof Vertex III, Långsele, Sweden) to the nearest 0.1 m at Alice Holt. Height measurements taken by Evans and Moustakas used a Laser Range Meter (Hilti PD40, Hilti, Schaan, Liechtenstein) to the nearest 0.1 m. The use of three different devices to assess height is likely to have increased measurement error in this parameter, at least if one was concerned with differences between the sites at which measurements were taken. A good test to determine the extent of this error would have been directly to compare measures of tree height taken using the three different instruments, unfortunately this was not possible. However, if a single measure of tree height is taken for each tree there are no significant difference in the measurements taken by the different instruments, once taxon and stage (adult or sapling) were taken into account (F = 5.43, N = 465, with eight taxa and 2 stages, P = 0.98).

*Diameter 10 cm above ground level ($D_{10}$)*

$D_{10}$ was measured for saplings in all three datasets: two measurements were made on each sapling using vernier callipers to the nearest 0.1 cm. The two measurements were taken, as far as practically possible, perpendicular to one other and averaged to produce one measurement per tree. A tape was not used to measure $D_{10}$ as vegetation and debris at the base of the trees made inserting a tape round the tree against its trunk extremely difficult to achieve in a consistent manner. As $D_{10}$ was not a repeated measure the point at which it was measured was not permanently marked as was DBH. The measurements were taken at a point that was determined to be 10cm above ground level (using vernier callipers).

*Crown radius (CRad)*

CRad was measured for adults in the ECN-W and ECN-AH datasets by visually projecting the crown margin onto the ground and measuring the two longest perpendicular diameters to the nearest cm using a measuring tape. The two measurements were halved and averaged to produce a single measurement per tree [14,20].

*Crown height (CH)*

CH was established for adults in the ECN-W and ECN-AH datasets by measuring the distance from the ground to the point where foliage occupied at least three of the four quadrants round the trunk [20], using a Laser Range Meter (Hilti PD40) to the nearest 0.1 m. These data were combined with height data for the same trees to estimate crown height (the distance between the top of the tree and the base of the crown), by subtracting the distance from the base of the crown from tree height [14,20].

*Light environment*

Light meter readings were taken on cloudy days (so that light arriving at the trees was as scattered as possible) during September 2012 for saplings in the ECN-W dataset, July-August 2013 for saplings in the ECN-AH dataset and September 2014 for a sample of 90 saplings in the OXF dataset [13]. We measured the percentage of incident light at the canopy reaching each tree ($L_c$) by measuring the light intensity beneath the canopy at three positions within 1 m of the trunk of each tree at a height of 1.3 m ($L_{cait}$ – the *i*th measurement of absolute light levels below the canopy taken at time t, where i = 1-3), and simultaneously in a large open gap nearby ($L_{oat}$ – the absolute light levels in the open at time t). To measure light levels we used two PAR Quantum sensors (SKP215, Skye Instruments Ltd, Llandrindod Wells, United Kingdom) that were both calibrated against the same reference lamp. The sensor used to measure light levels under the canopy was used in conjunction with a meter (SKP 200, Skye Instruments Ltd, Llandrindod Wells, United Kingdom) recording to one decimal place; the one measuring light levels in the open was used with a datalogger (SDL5050 DataHog 2, Skye Instruments Ltd, Llandrindod Wells, United Kingdom). Measurements from the sensor in the open gap were made every 10 s with the mean of these more frequent values recorded every 10 min. We calculated three light intensity values for each tree, which are the proportion of the available light that reached each tree's position ($L_{ci}$, where i = 1-3):

$$L_{ci} = L_{cait}/L_{oat} \tag{1}$$

$L_{c1}$, $L_{c2}$ and $L_{c3}$ for each tree were averaged to produce a single value ($L_c$) for each individual tree.

*Canopy openness*

This light transmission coefficient is estimated using fish-eye-lens photographs taken under canopies that are dominated by a single taxon. The fish-eye-lens photographs are taken at 1.35 m above the ground and orientated to magnetic North. The percentage of canopy openness was analysed for individual circular sections of canopy using Gap Light Analyzer software (http://www.ecostudies.org/gla/), following the method described in [20]. The gap light analyser software allows the crown of a tree to be identified in the photograph by the operator and then estimates the percentage of canopy openness for circular sections of the crown. The degree of canopy openness depends on the structure of the crown and the size and shape of the leaves, this variable is used in SORTIE to filter out light hitting the canopy and so modify the light environment below the tree. Differences in canopy openness and canopy dimensions between taxa create a patchy light environment in the forest.

**Summary Variables**

We generated the taxon-specific summary statistics relating to the allometry and growth equations required by SORTIE [8]. These are:

*Allometry*

Taxon-specific allometric functions describe the tree's size and shape.

<u>Saplings (trees with DBH<10 cm)</u>

To describe the allometry of saplings, two relationships are used – a linear one between $D_{10}$ (trunk diameter at 10 cm above ground level) and DBH, and a power function between $D_{10}$ and height (H).

$$DBH = a + bD_{10} \tag{2}$$

$$H = aD_{10}^b \text{ or } \log H = \log a + b \log D_{10} \tag{3}$$

<u>Adults (trees with DBH>10 cm)</u>

To adequately describe the size and shape of adult trees requires three allometric relationships to be parameterised, power relationships between crown radius (CRad) and DBH, crown height (CH) and tree height; and an exponential relationship between height and DBH, with an asymptote at maxH.

$$CRad = aDBH^b, \text{ or } \log CRad = \log a + b \log DBH \tag{4}$$

$$CH = aH^b, \text{ or } \log CH = \log a + b \log H \tag{5}$$

$$H = 1.35 + (maxH - 1.35)(1 - e^{-bDBH}) \tag{6}$$

*Growth*

<u>Saplings</u>

Radial growth is assumed to be described by a Michaelis-Menten function that relates growth in DBH ($G_{sap}$, in cm yr$^{-1}$) to light availability (L, expressed as a percentage of daylight), combined with a power function of the effect of size. Michaelis-Menten functions are specific forms of dose-response curves where the rate of a response variable depends on the concentration of a substrate. Here sapling growth is the response variable and the intensity of light is the substrate on which growth depends [13].

$$G_{sap} = \left(\alpha L / \left(L + (\alpha / \beta)\right)\right) D_{10}^{\phi} \quad (7)$$

$\alpha$ is the asymptotic growth a high light levels, $\beta$ is the slope of the growth function at zero light. $D_{10}^{\phi}$ is the size effect to determine the most appropriate value of $\phi$ we fitted models with $\phi$ = 0 - 1, and report the best fitting model (as determined by the lowest residual standard error) which was $\phi$ = 0.845 (which gave a residual standard error of 0.005 with 116 degrees of freedom).

Adults

Adult radial growth rate was assumed to be related to maximum radial growth rate that a taxon can attain devalued by a size effect, so that in general trees grow more slowly as they get larger.

$$G_{adult} = MaxG \times SE \quad (8)$$

The size effect SE is given by:

$$SE = e^{-0.5(\ln(DBH/x_0)/x_b)^2} \quad (9)$$

$x_0$ and $x_b$ are estimated parameters.

**Data analysis**

As both the dependent and the independent variables were subject to sampling error, ranged major axis (RMA) model II regression [21] was used to analyse the relationships between sapling $D_{10}$ and height (equation 3), sapling $D_{10}$ and DBH (equation 2), adult CRad and DBH (equation 4), and adult CH and height (equation 5). We used the lmodel2 procedure in the lmodel2 library [22] implemented in R 2.15.2 (Ref. [23]). As we had longitudinal data on both adult height and DBH (equation 6) we used repeated measures ANOVA with DBH as the independent variable and height as the dependent variable and individual code as a random effect to avoid pseudo-replication of trees that had been measured more than once. For this analysis we used the lmer procedure in the lme4 library [24] in R 2.15.2 (Ref. [23]). To analyse the relationship between sapling growth rates and light (equation 7) we used the MM2 procedure in the drc library [25] of R to fit a two parameter Michaelis-Menten function to the relationship between the growth rate and the light environment of individual saplings.

Equation 9 is a two-parameter ($x_0$ and $x_b$) negative exponential distribution. In order to estimate $x_0$ and $x_b$ inverse modelling was employed (identifying the parameters of a distribution from data). Maximum likelihood estimation was used for fitting the two parameters of the negative exponential distribution [26] using data on adult tree growth rates in ECN-W and ECN-AH.

## Data Records

The data contained in this data descriptor have been deposited in Datadryad (http://dx.doi:10.5061/dryad.2c1s7). All data include codes to identify the individual trees: for ECN-W these are derived by adding (tree number) to (plot number x 100); for ECH-AH they are derived by adding (cell identity code) to (plot identity code x 100); for OXF all trees have individual coded tags and these numbers were used as the identity codes. Individual codes can be used to identify individual trees within a given dataset but may be replicated between datasets. Our study is primarily concerned with allometric relationships of saplings and adult trees. We also provide the original data needed to derive these data. The majority of the DBH and height data are publicly available: All ECN data used here (DBH and height data for datasets ECN-W and ECN-AH) are available on request from Centre for Ecology and Hydrology (http://data.ecn.ac.uk/access.asp); the DBH records associated with dataset OXF have been published at http://ctfs.arnarb.harvard.edu/Public/plotdataaccess/index.php from where they can be freely downloaded.

**Sapling allometry DBH, $D_{10}$ & Height – data record 1**

Contains data on DBH (cm), $D_{10}$ (cm) and Height (m) for a total of 145 saplings for the eight taxa under consideration. Data are drawn from three datasets (ECN-W, ECN-AH and OXF). The year in which the DBH and height and $D_{10}$ data were recorded is reported for each individual. Data record 1 is stored as a tab delimited text file (Data citation 1), and is available from the Dryad Digital Repository, an up-to-date file is maintained at www.predictivecology.com. The dataset was last updated October 16 2014.

**Adult allometry DBH, Height & Crown height – data record 2**

Contains data on DBH (cm), Height (m), Crown height (m) and Crown radius (m) for a total of 297 adult trees for the eight taxa under consideration. Data are drawn from two datasets (ECN-W and ECN-AH). The year in which DBH, height and crown height and radius were recorded are reported for each individual. Data record 2 is stored as a tab delimited text file (Data citation 1), and is available from the Dryad Digital Repository, an up-to-date file is maintained at www.predictivecology.com. The dataset was last updated October 16 2014.

**All trees height v DBH – data record 3**

Contains data on DBH (cm) and Height (m) for 481 individuals for the eight taxa under consideration. Data are drawn from two datasets (ECN-W and ECN-AH). Repeated measures on each individual results in 1211 records, the year of each measurement is reported. Data record 3 is stored as a tab delimited text file (Data citation 1), and is available from the Dryad Digital Repository, an up-to-date file is maintained at www.predictivecology.com. The dataset was last updated February 5 2015.

**Sapling growth – data record 4**

Contains data on DBH growth rates (cm yr$^{-1}$) for the periods between measurements, the mean growth rate, $D_{10}$ (cm), and the fraction of ambient light in the tree's environment for 129 individuals representing seven of the eight taxa under consideration to parameterise equation 7. Data are drawn from two datasets (ECN-W and OXF). The year in which $D_{10}$ and light were measured is reported. Data record 4 is stored as a tab delimited text file (Data citation 1), and is available from the Dryad Digital Repository, an up-to-date file is maintained at www.predictivecology.com. The dataset was last updated October 16 2014.

**All trees growth – data record 5**

Contains data on DBH growth rates (cm yr$^{-1}$) for both adults and saplings for the periods between measurements and the mean growth rate for 439 individuals of the eight taxa under consideration. Data are drawn from three datasets (ECN-W, OXF and ECN-AH). Data record 5 is stored as a tab delimited text file (Data citation 1), and is available from the Dryad Digital Repository, an up-to-date file is maintained at www.predictivecology.com. The dataset was last updated February 5 2015.

**Canopy openness – data record 6**

Contains data on canopy openness for 165 single taxon stands of the eight taxa under consideration. Data record 6 is stored as a tab delimited text file (Data citation 1), and is available from the Dryad Digital Repository, an up-to-date file is maintained at www.predictivecology.com. The dataset was last updated October 16 2014.

**SORTIE parameter file – data record 7**

Contains data on 16 parameters for each of the eight taxa considered here. These allow the instantiation of the equations 1-8 listed above. In conjunction with parameters on mortality and dispersal they also allow SORTIE to be run to produce projections of United Kingdom lowland woodlands. Data record 7 is stored as a tab delimited text file (Data citation 1), and is available from the Dryad Digital Repository, an up-to-date file is maintained at www.predictivecology.com. The dataset was last updated February 5 2015.

## Technical Validation

Once we had compiled data into the collated files, data entries were completed and verified using a number of techniques:

1. Any missing data were checked by examining the original data files obtained from ECN or Malhi and Butt and field notebooks
2. Taxonomic codes were standardised and checked by counting the frequency with which each code appeared, examining any which were represented by few entries, and correcting any typographical errors that were revealed
3. Maxima, minima, means and variances were calculated for all variables and outliers, and checked against original data records
4. Each file was created from the original data twice separated by at least one month, the sequence of data in at least one column per dataset was used as an index variable, and the order obtained in the two datasets compared against each other. Any discrepancies were checked against the original datasets
5. We have plotted the summary parameters in data record 7, to determine whether the predicted relationships are reasonable and in accordance with the most complete set of similar relationships found in [14]. These can be seen in Figs. 2 and 3, and will be updated along with data record 7 and new versions posted at www.predictivecology.com.

## Acknowledgements
We thank the University of Oxford for access to Wytham Woods and Forest Research for access to Alice Holt. Michelle Taylor and Mike Morecroft collected much of the original ECN data at Wytham Woods.

## Author Contributions

MRE collected data on $D_{10}$, crown height, crown radius, light environment around trees, conducted the statistical analyses, collated the datasets and drafted the manuscript

AM collected data on D10, crown height, crown radius, light environment around trees, calculated $x_0$ and $x_b$ and drafted the manuscript

GC collected data on canopy openness, and commented on the manuscript

YM and NB collected data in dataset OXF on DBH and height, NB commented on the manuscript

DP and SS collected data in dataset ECN-W, and commented on the manuscript

SB collected data in dataset EN-AH, and commented on the manuscript

## Competing Financial Interests

The authors have no conflicting financial interests

# References

1   Medawar, P. *The Limits Of Science.*  (Oxford University Press, 1984).
2   Lonergan, M. Data availability constrains model complexity, generality, and utility: a response to Evans et al. *Trends Ecol. Evol.* **29,** 301-302 (2014).
3   Evans, M. R. *et al.* Data availability and model complexity, generality, and utility: a reply to Lonergan. *Trends Ecol. Evol.* **29,** 302-303 (2014).
4   Moustakas, A. *et al.* Long-term mortality patterns of the deep-rooted Acacia erioloba: The middle class shall die! *J. Veg. Sci.* **17,** 473-480 (2006).
5   Evans, M. R. *et al.* Predictive systems ecology. *Proc. R. Soc. Ser. B* **280,** 20131452 (2013).
6   Evans, M. R. *et al.* Do simple models lead to generality in ecology? *TREE* **28,** 578-583 (2013).
7   DEFRA. *The UK National Ecosystem Assessment: Synthesis Of The Key Findings.*, (Department for the Environment, Food and Rural Affairs, 2011).
8   Pacala, S. W. *et al.* Forest models defined by field measurements: estimation, error analysis and dynamics. *Ecol. Monogr.* **66,** 1-43 (1996).
9   Purves, D. & Pacala, S. Predictive models of forest dynamics. *Science* **320,** 1452-1453 (2008).
10  Strigul, N., Pristinski, D., Purves, D., Dushoff, J. & Pacala, S. Scaling from trees to forests: tractable macroscopic equations for forest dynamics. *Ecol. Monogr.* **78,** 523-545 (2008).
11  Purves, D. W., Lichstein, J. W., Strigul, N. & Pacala, S. W. Predicting and understanding forest dynamics using a simple tractable model. *Proc. Natl. Acad. Sci. USA* **105,** 17018-17022 (2008).
12  Kunstler, G., Coomes, D. A. & Canham, C. D. Size-dependence of growth and mortality influence the shade tolerance of trees in a lowland temperate rain forest. *J. Ecol.* **97,** 685-695 (2009).
13  Coomes, D. A., Kunstler, G., Canham, C. D. & Wright, E. A greater range of shade-tolerance niches in nutrient-rich forests: an explanation for positive richness–productivity relationships? *J. Ecology* **97,** 705-717 (2009).
14  Kunstler, G., Allen, R. B., Coomes, D. A., Canham, C. D. & Wright, E. F. *SORTIE/NZ Model Development.*  55 (Landcare Research New Zealand Ltd, 2011).
15  Schmidt-Nielsen, K. *Scaling: Why Is Animal Size So Important?* , (Cambridge University Press, 1984).
16  Bonner, J. T. *Why Size Matters: From Bacteria To Blue Whales.*  (Princeton University Press, 2006).
17  Mackie, E. D. & Matthews, R. W. *Forest Mensuration, A Handbook For Practitioners.*, (HMSO, 2006).
18  Sykes, J. M. & Lane, A. M. J. *The United Kingdom Environmental Change Network: Protocols For Standard Measurements At Terrestrial Sites.*,  (The Stationery Office, 1996).
19  Moustakas, A. & Evans, M. R. Effects of growth rate, size, and light availability on tree survival across life stages: a demographic analysis accounting for missing values and small sample sizes. *BMC Ecology* (2015).
20  Canham, C. D., Coates, K. D., Bartemucci, P. & Quaglia, S. Measurement and modeling of spatially-explicit variation in light transmission through interior cedar-hemlock forests of British Columbia. *Can. J. For. Res.* **29,** 1775-1783 (1999).
21  Sokal, R. R. & Rohlf, F. J. *Biometry - The Principles And Practice Of Statistics In Biological Research.* 3rd edn,  (W.H. Freeman, 1995).
22  Legendre, P. lmodel2: Model II Regression. R package version 1.7-0.  (2011).

23    R Development Core Team. *R; A language and environment for statistical computing.*, (R Foundation for Statistical Computing, 2008).
24    lme4: Linear mixed-effects models using S4 classes. R package version 0.999999-0. (2012).
25    Ritz, C. & Streibig, J. C. Bioassay analysis using R. *J. Statist. Software* **12** (2005).
26    Harris, C. M. & Sykes, E. A. Likelihood estimation for generalized mixed exponential distributions. *Naval Research Logistics* **34,** 251-279 (1987).

## Data Citations
1    Evans, M.R., Moustakas, A., Carey, G., Malhi, Y., Butt, N., Benham, S., Pallett, D., & Schäffer, S. *Dryad* http://dx.doi:10.5061/dryad.2c1s7

**Table 1.** List of taxa included in the data files, including the taxonomic code used in all data files, Latin name and common name of each taxon.

| Latin name | Common name | Taxonomic code |
|---|---|---|
| *Acer pseudoplatanus* | Sycamore | ACERPS |
| *Fraxinus excelsior* | European ash | FRAXEX |
| *Quercus robur* | Pedunculate oak | QUERRO |
| *Fagus sylvatica* | European beech | FAGUSY |
| *Corylus avellana* | Common hazel | CORYAV |
| *Crataegus monogyna* | Common hawthorn | CRATMO |
| *Acer campestre* | Field maple | ACERCA |
| *Betula* spp. | Birch | BETUSP |

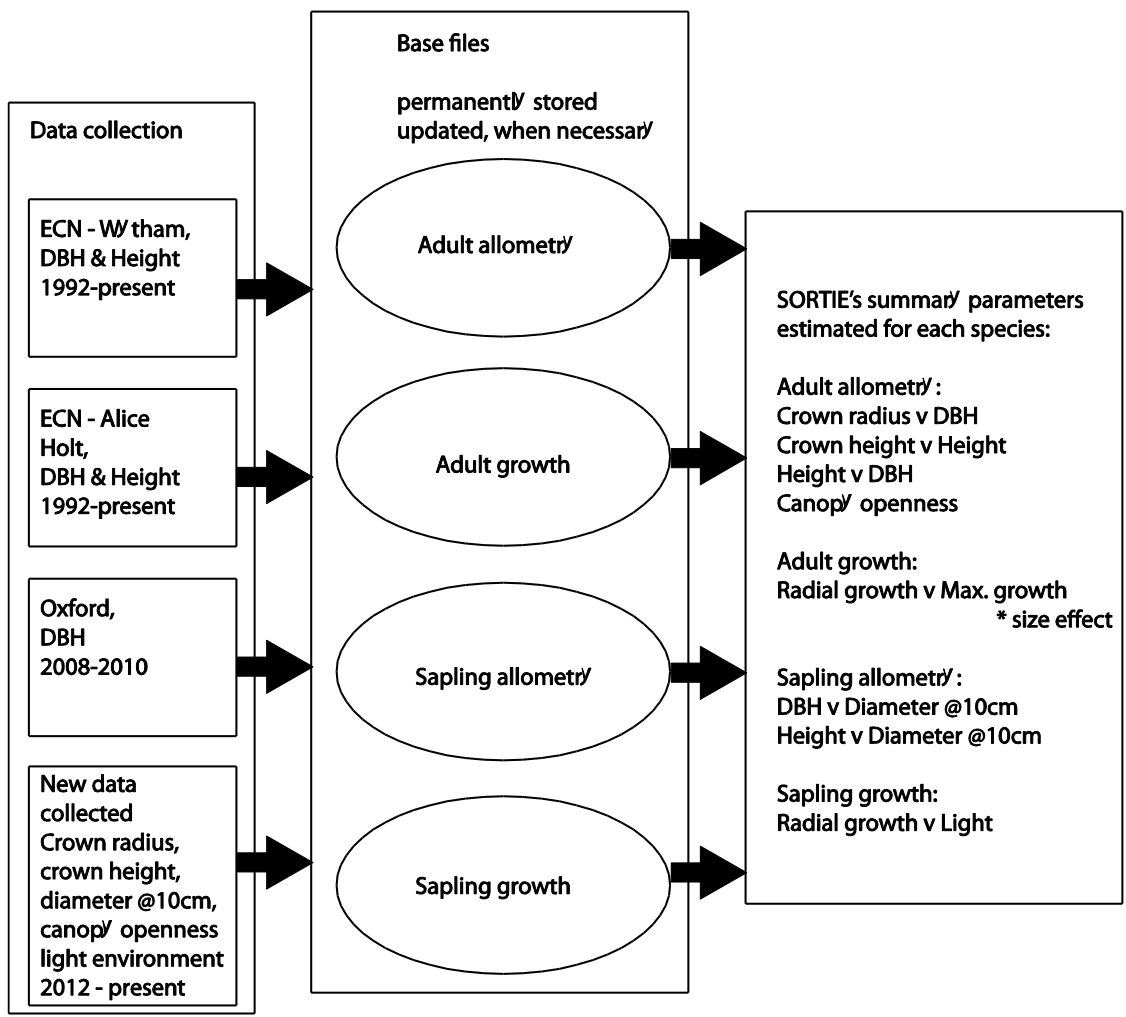

**Figure 1.** Workflow for database construction and parameter estimation. Base files and files containing SORTIE's parameters are contained in this database release. Both types of files can be refreshed when new data are included by re-running analyses on updated base files.

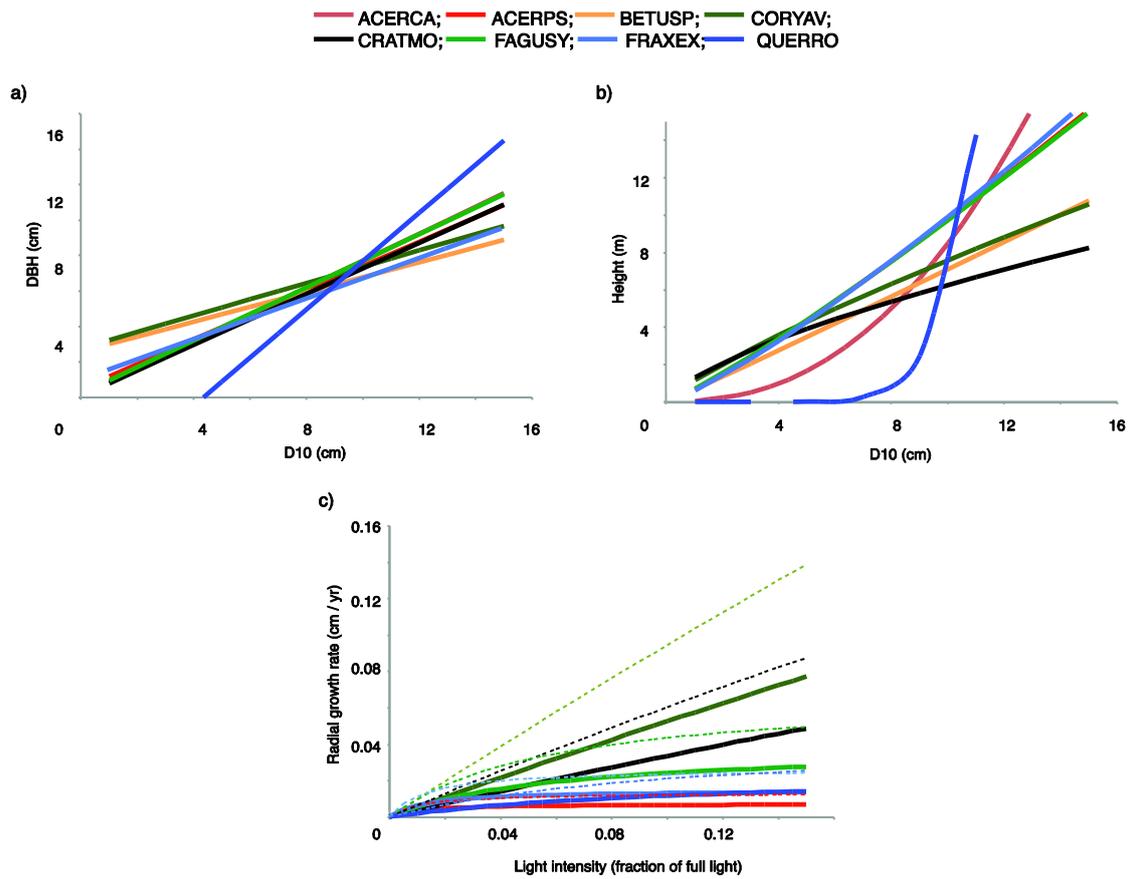

**Figure 2.** Allometric relationships and growth rates for saplings trees.

a) Allometric relationships between $D_{10}$ and DBH for the eight tree taxa considered here, estimated using parameters in data record 7 in equation 2.

b) Allometric relationships between $D_{10}$ and height for the eight tree taxa considered here, estimated using the parameters in data record 7 in equation 3.

c) Allometric growth in different light environments for the six taxa for which the relevant parameters could be estimated. Functions estimated using parameters in data record 7 in equation 7. Solid lines are for 5cm DBH trees, dotted lines for 10cm DBH trees.

ACERCA – Field Maple (*Acer campestre*); ACERPS – Sycamore (*Acer pseudoplatanus*); BETUSP – Birch (*Betula* spp.); CORYAV – Hazel (*Corylus avellana*); FAGUSY – Beech (*Fagus sylvatica*); FRAXEX – Ash (*Fraxinus excelsior*), QUERRO – Pedunculate Oak (*Quercus robur*).

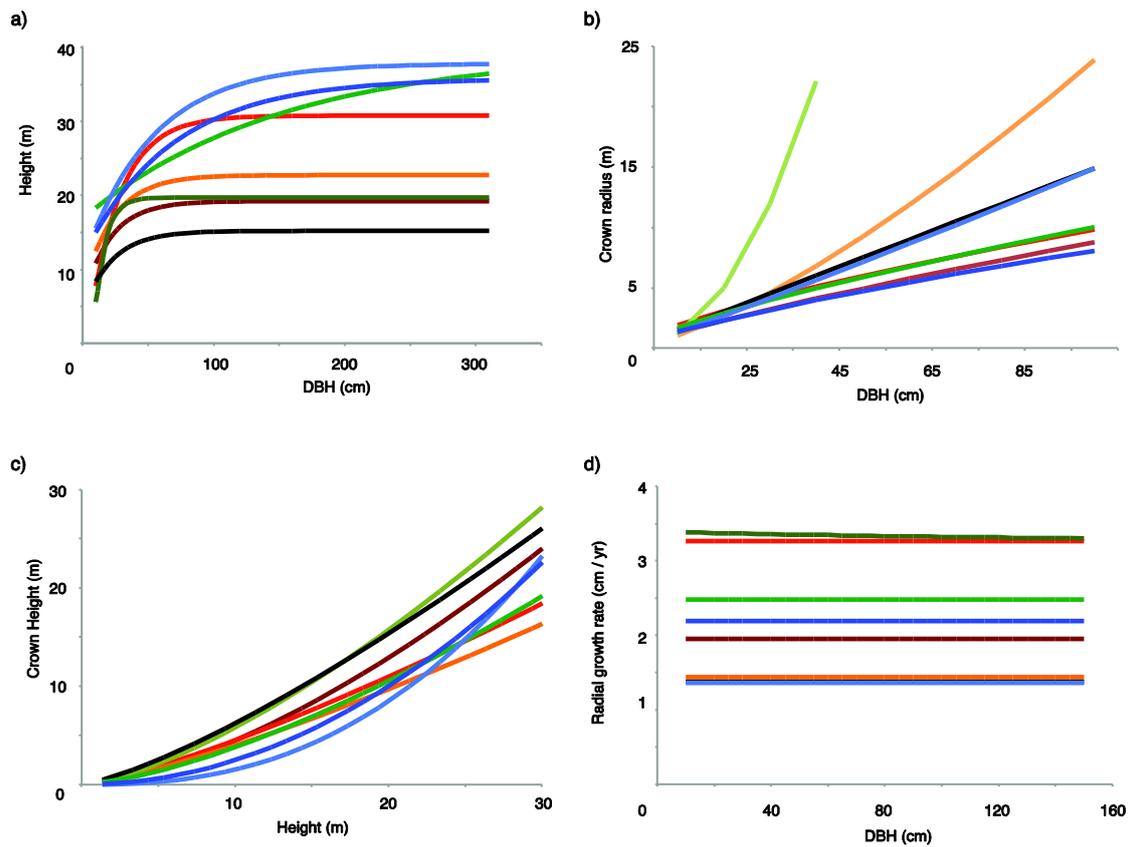

**Figure 3.** Allometric relationships and growth rates for adult trees.

a) Allometric relationships between DBH and height for the eight tree taxa considered here, estimated using parameters in data record 7 in equation 6.

b) Allometric relationships between DBH and crown radius for the eight tree taxa considered here, estimated using parameters in data record 7 in equation 4.

c) Allometric relationships between crown height and height for the eight tree taxa considered here, estimated using parameters in data record 7 in equation 5.

d) Allometric annual diameter growth rate for each of the eight taxa considered here, estimated using parameters in data record 7 in equation 8. For legend see figure 2.